\documentclass[12pt]{article}
\usepackage{epsfig}
\begin{document}
\begin{center}
{\Large {\bf A superstructure over the Farhi - Susskind Technicolor model}

\vskip-40mm \rightline{\small ITEP-LAT/2009-16 } \vskip 30mm

{
\vspace{1cm}
{ M.A.~Zubkov }\\
\vspace{.5cm} {\it  ITEP, B.Cheremushkinskaya 25, Moscow, 117259, Russia }
\\{\it  Moscow Institute of Physics and Technology, 141700,
Dolgoprudnyi, Moscow Region, Russia } }}
\end{center}

\begin{abstract}
We suggest the model with the gauge group $ ... \otimes SU(6) \otimes
SU(5)\otimes SU(4) \otimes SU(3) \otimes SU(2) \otimes U(1)$. This group is the
infinite continuation of the gauge group $SU(4) \otimes SU(3) \otimes SU(2)
\otimes U(1)$ of Farhi - Susskind model. The constructed model contains
fermions from the fundamental representations of any $SU(N)$ subgroups of the
gauge group. In the construction of the model we use essentially the
requirement that it posseses an additional discrete symmetry $\cal Z$ that is
the continuation of the $Z_6$ symmetry of the Standard Model. It has been found
that there exists such a choice of the hypercharges of the fermions that the
chiral anomaly is absent while the symmetry $\cal Z$ is preserved.

\end{abstract}

\section{Introduction}

Recently we have shown \cite{Z2007t} that the $Z_6$ symmetry of the Standard
Model \cite{Z6,Z6_,Z6__,BVZ2003,Z2007} (SM) can be continued to the Technicolor
models (TC). It was shown that among various models only a few ones possess the
new discrete symmetry $\cal Z$. In particular, for the Farhi - Susskind model
with Technicolor group $SU(N_{TC})$ there are two possibilities: $N_{TC} = 2$
and $N_{TC} = 4$. It is worth mentioning that the $SU(2)$  Farhi - Susskind
model \cite{FS} suffers from the vacuum alignment problems. That's why we do
not consider it as realistic and the only possibility remains that is the
$SU(4)$ Farhi - Susskind model. This model (together with the Standard Model)
has the gauge group $SU(4) \otimes SU(3) \otimes SU(2) \otimes U(1)$. The
hypercharge assignment for the technifermions is fixed by the additional
discrete symmetry up to an integer number.

The structure of the gauge group $SU(4) \otimes SU(3) \otimes SU(2) \otimes
U(1)$ prompts its possible continuation as the infinite sequence of $SU(N)$
subgroups:
\begin{equation}
G = ... \otimes SU(6)\otimes SU(5)\otimes SU(4) \otimes SU(3) \otimes SU(2)
\otimes U(1)/{\cal Z}, \label{G}
\end{equation}
where $\cal Z$ is the discrete group to be specified below.

The question arises: is this possible to continue the $Z_6$ symmetry of the
Standard Model to this sequence. In the present paper we construct such a
continuation. We arrange the fermions of the model in the fundamental
representations of $SU(N)$ subgroups of (\ref{G}). In general case the model
with the gauge group (\ref{G}) suffers from chiral anomalies of different
types. It is not obvious a priori that there exists the hypercharge assignment
of the fermions such that the chiral anomalies are absent while the additional
discrete symmetry is preserved. Below we show that it is possible to satisfy
both requirements simultaneously.

The Technicolor interaction alone serves only as a source of Electroweak gauge
symmetry breaking. Usually in order to make Standard Model fermions massive
extra gauge interaction is added, which is called Extended Technicolor (ETC)
\cite{Technicolor,Technicolor_,Technicolor__,ExtendedTechnicolor,ExtendedTechnicolor_,ExtendedTechnicolor__,ExtendedTechnicolor___,ExtendedTechnicolor____}.
In this gauge theory the Standard Model fermions and technifermions enter the
same representation of the Extended Technicolor group. Standard Model fermions
become massive because they may be transformed into technifermions with
ejecting of the new massive gauge bosons. The ETC models suffer from extremely
large flavor - changing amplitudes and unphysically large contributions to the
Electroweak polarization operators
\cite{Technicolor,Technicolor_,Technicolor__}. The possible way to overcome
these problems is related to the behavior of chiral gauge theories at large
number of fermions or for the higher order representations
\cite{minimal_walking,minimal_walking_}.

In the present paper we do not concretize the mechanism of fermion mass
generation. It may be either of  the ETC type or some unknown mechanism related
to a higher energy scale. Whatever mechanism is it defines (together with the
chiral symmetry breaking) the correspondence between the left handed and the
right - handed fermions that is identified with parity conjugation of spinors.
Our model as well as the SM and the Farhi - Susskind model contains left handed
doublets and right handed singlets.  That's why the mentioned correspondence
connects pairs of right - handed singlets with the components of certain left -
handed doublets. In particular, the component of the left - handed doublet that
corresponds to the right - handed electron is called left - handed electron.
Let us denote the left - handed doublet of the 1-st generation SM leptons by
$\Theta$. The given correspondence can be written as: $\Omega^1_{i}\Theta^{i} =
\nu_L;\, \Omega^2_{i}\Theta^{i} =  e^-_L$, where we introduce the auxiliary
field $\Omega \in SU(2)$. The same field $\Omega$ applied to the other left -
handed doublets gives the parity conjugated partners to the remaining right -
handed singlets. The physical quantities do not depend on $\Omega$ and it can
always be chosen equal to unity. This choice corresponds to the unitary gauge.
The mass term generated via the ETC (or other) interactions gives the
transition amplitudes between the right - handed singlets and their parity
partners. At the same time the dynamical fermion terms contain mixing. The
requirement that $\Omega$ is the same (up to the gauge transformation) for all
left - handed doublets is necessary for the correct realization of the
Electroweak symmetry breakdown. It is worth mentioning that in the $SU(4)$
Farhi - Susskind model with ETC interaction considered as a perturbation the
vacuum alignment works properly and leads to the correct Electroweak symmetry
breaking \cite{Align}. We also notice that the field $\Omega$ is not dynamical,
i.e. there is no integration over $\Omega$ in the functional integral that
defines correlation functions. Fixing of $\Omega$ means that one of the
equivalent vacua is chosen during the spontaneous breakdown of Electroweak
symmetry.

Extra $SU(N)$ ($N>4$) gauge interactions present in (\ref{G}) may be observed
in principle at the energies above the Technicolor scale. We briefly concern
their properties at the end of the paper. Throughout the paper we call $SU(N)$
subgroups for $N>4$ the Hypercolor groups. We also feel it appropriate to refer
to the sequence (\ref{G}) as to the Hypercolor tower.

\section{The model}

In our approach the theory contains $U(1)$ gauge group and the groups $SU(N)$
with any $N$. So, the gauge group of the theory is (\ref{G}). Next, we suppose,
that in the theory any fermions are present that belong to the fundamental
representations of the $SU(N)$ subgroups of $G$. So, the possible fermions are
right - handed $\Psi_{A, Y}^{\alpha i_{k_N} ... i_{k_3} i_{k_2}}$ and left -
handed $\Theta_{\dot{\beta} A, Y}^{ i_{k_N} ... i_{k_3} i_{k_2}}$, where
$\alpha$ and $\dot{\beta}$ are spinor indices, $A$ enumerates generations while
index $i_k$ belongs to the subgroup $SU(k)$. Here $Y$ is the $U(1)$ charge of
the given fermion. In particular, the fermions $\Psi_{A; Y}$ are present that
have no indices and the only subgroup that acts on $\Psi_{A; Y}$ is $U(1)$.
Moreover, we suppose that the fermions are present such that $G$ does not act
on them at all. We denote them $\Psi_{A;0}$. All fermions in the theory are two
- component spinors. We also suppose from the very beginning that the $SU(2)$
group acts on the left - handed spinors only. The action of parity conjugation
on them will be considered later. For the simplicity we omit below both spinor
and generation indices. So, our fermions are

\begin{eqnarray}
U(1): && \Psi_0, \Psi_{Y_{1}}, \Psi_{Y^{\prime}_{1}},...;\nonumber\\
U(1), SU(2): && \Theta^{i_2}_{Y_2}, \Theta^{i_2}_{Y^{\prime}_2}, ...; \nonumber\\
U(1), SU(3): && \Psi^{i_3}_{Y_3}; \Psi^{i_3}_{Y^{\prime}_3}, ...; \nonumber\\
U(1), SU(2), SU(3): && \Theta^{i_3 i_2}_{Y_{32}},\Theta^{i_3 i_2}_{Y^{\prime}_{32}},... ;\nonumber\\
U(1), SU(4): && \Psi^{i_4}_{Y_4},\Psi^{i_4}_{Y^{\prime}_4},...;\nonumber\\
U(1), SU(2), SU(4): && \Theta^{i_4 i_2}_{Y_{42}},\Theta^{i_4 i_2}_{Y^{\prime}_{42}},...;\nonumber\\
U(1), SU(3), SU(4): && \Psi^{i_4 i_3}_{Y_{43}}, \Psi^{i_4 i_3}_{Y^{\prime}_{43}},...;\nonumber\\
U(1), SU(2),SU(3),SU(4): &&  \Theta^{i_4 i_3 i_2}_{Y_{432}},\Theta^{i_4 i_3
i_2}_{Y^{\prime}_{432}},...; \nonumber\\
&&...\label{F}
\end{eqnarray}
Here in each row we list the subgroups of $G$ that act on the fermions listed
in the row. In each row the allowed values of $U(1)$ charge are denoted by $Y,
Y^{\prime}$, etc.

Let us consider the first row. Here in order to reproduce the Standard Model we
restrict ourselves by the values of $Y$ equal to $0$ and $-2$. Next, the second
row must contain the only element with $Y = -1$. The third row contains two
elements with $Y = \frac{4}{3}$ and $Y = -\frac{2}{3}$. In the forth row we
have the only element with $Y = \frac{1}{3}$. This row completes the Standard
Model and we enter the rows related to its ultraviolet completion.

Now let us consider the second four rows in (\ref{F}). We suggest them in the
form that represents $SU(4)$ Farhi - Susskind model of Technicolor \cite{FS}.
In \cite{Z2007t} we have derived the hypercharge assignment for the
technifermions such that the chiral anomaly is absent while the additional
discrete symmetry is preserved. As a result the hypercharge assignment is the
following. In the $5$ - th row there are two elements with $Y_4 = \frac{1}{2}-
6K + 1$ and $Y^\prime_4 = \frac{1}{2}- 6K -1$ (were $K$ is an arbitrary integer
number). In the $6$ -th row we have the only element with $Y_{42} =
\frac{1}{2}- 6K$, where $K$ is the same as in the previous row. In the $7$ - th
row there are two elements with $Y_{43} = -\frac{\frac{1}{2}- 6K}{3} + 1$ and
$Y^\prime_{43} = -\frac{\frac{1}{2}- 6K}{3} -1$. The $8$ -th row contains the
only element with $Y_{432} = -\frac{\frac{1}{2}- 6K}{3}$. Again, in these two
rows $K$ is the same as before.

Let us specify how parity conjugation $\cal P$ acts on the fermions. If only
two fermions $\chi^{\alpha}$ and $\eta_{\dot{\alpha}}$ are present, then ${\cal
P}\chi^{\alpha}(t,\bar{r}) = i \eta_{\dot{\alpha}}(t,-\bar{r}); {\cal P}
\eta_{\dot{\alpha}}(t,\bar{r})= i \chi^{\alpha}(t,-\bar{r})$. In our case we
require that for any configuration of $SU(N)$ ($N > 2$) indices there exist two
right - handed spinors and one $SU(2)$ doublet. The parity conjugation connects
each of the right handed spinors with a component of the $SU(2)$ doublet. Thus
\begin{eqnarray}
&& {\cal P}\Psi_0(t,\bar{r}) = i \Omega^1_{i_2}(t,-\bar{r})
\Theta^{i_2}_{-1}(t,-\bar{r}); {\cal P}\Psi_{-2} = i
\Omega^2_{i_2}\Theta^{i_2}_{-1};\nonumber\\
&& {\cal P}\Psi^{i_3}_{\frac{4}{3}} = i \Omega^1_{i_2} \Theta^{i_3
i_2}_{\frac{1}{3}}; {\cal P}\Psi^{i_3}_{-\frac{2}{3}} = i
\Omega^2_{i_2}\Theta^{i_3 i_2}_{\frac{1}{3}};\nonumber\\
&& {\cal P}\Psi^{i_4}_{Y_{4}} = i \Omega^1_{i_2} \Theta^{i_4 i_2}_{Y_{42}};
{\cal P}\Psi^{i_4}_{Y^\prime_{4}} = i
\Omega^2_{i_2}\Theta^{i_4 i_2}_{Y_{42}};\nonumber\\
&& {\cal P}\Psi^{i_4 i_3}_{Y_{43}} = i \Omega^1_{i_2} \Theta^{i_4
i_2}_{Y_{432}}; {\cal P}\Psi^{i_4 i_3}_{Y^\prime_{43}} = i
\Omega^2_{i_2}\Theta^{i_4 i_3 i_2}_{Y_{432}};\nonumber\\
&& ... \label{P}
\end{eqnarray}

Here $\Omega$  is the auxiliary $SU(2)$ field mentioned in the Introduction.
The physical sense of the field $\Omega$ is that it peeks up the parity partner
for each right - handed spinor.

The correspondence between our notations and the conventional ones is the
following (we list here the case $K = 0$ for the first generation only):
\begin{eqnarray}
 && \Psi_0 = \nu_R; \Psi_{-2} = e^-_R; {\cal P}\Psi_0(t,\bar{r}) = i \nu_L(t,-\bar{r}); {\cal P}\Psi_{-2} = i e^-_L ;\nonumber\\
 && \Psi^{i_3}_{\frac{4}{3}} = u_R; \Psi^{i_3}_{-\frac{2}{3}} = d_R; {\cal P}\Psi^{i_3}_{\frac{4}{3}} = i
 u_L;  {\cal P}\Psi^{i_3}_{-\frac{2}{3}} = i d_L;\nonumber\\
 && \Psi^{i_4}_{\frac{3}{2}} = N_R; \Psi^{i_4}_{-\frac{1}{2}}=E_R; {\cal P}\Psi^{i_4}_{\frac{3}{2}} = i
 N_L;  {\cal P}\Psi^{i_4}_{-\frac{1}{2}} = i E_L ;\nonumber\\
 && \Psi^{i_4 i_3}_{\frac{5}{6}} = U_R; \Psi^{i_4 i_3}_{-\frac{7}{6}}= D_R;
 {\cal P}\Psi^{i_4 i_3}_{\frac{5}{6}} = i U_L; {\cal P}\Psi^{i_4 i_3}_{-\frac{7}{6}} =
 i D_L.\label{i}
\end{eqnarray}

 It is worth
mentioning that the fermions of the first generation listed here do not
diagonalize the mass matrix. Instead the certain linear combinations of the
listed fermions diagonalize the mass matrix thus giving rise to mixing angles
and flavor changing amplitudes.

Before dealing with the next rows let us remind what we call the additional
$Z_6$ symmetry in the Standard Model and how can it be continued to the
Hypercolor interactions.

\section{$\cal Z$ symmetry}

 Within the Standard Model for any path $\cal C$,
we may calculate the elementary parallel transporters $\Gamma = {\rm P} \, {\rm
exp} (i\int_{\cal C} C^{\mu} dx^{\mu}), U = {\rm P} \, {\rm exp} (i\int_{\cal
C} A^{\mu} dx^{\mu}), e^{i\theta} = {\rm exp} (i\int_{\cal C} B^{\mu}
dx^{\mu})$, where $C$, $A$, and $B$ are correspondingly $SU(3)$, $SU(2)$ and
$U(1)$ gauge fields of the Standard Model. The parallel transporter
correspondent to each fermion of the Standard Model is the product of the
elementary ones listed above. Therefore,  the elementary parallel transporters
are encountered in the theory only in the following combinations:
$e^{-2i\theta};\,U\, e^{-i\theta};\Gamma \, U \, e^{ \frac{i}{3} \theta};
\Gamma \, e^{ -\frac{2i}{3} \theta}; \Gamma \, e^{ \frac{4i}{3} \theta}$. It
can be easily seen \cite{BVZ2003} that {\it all} the listed combinations are
invariant under the following $Z_6$ transformations: $U \rightarrow  U e^{i\pi
N}, \theta  \rightarrow  \theta +  \pi N,  \Gamma \rightarrow  \Gamma e^{(2\pi
 i/3)N}$, where $N$ is an arbitrary integer number.  This symmetry allows to define the
Standard Model with the gauge group $SU(3)\times SU(2) \times U(1)/{Z_6}$
instead of the usual $SU(3)\times SU(2) \times U(1)$.

In \cite{Z2007t} we have suggested the way to continue this symmetry to the
Technicolor extension of the Standard Model. Now we generalize the construction
of \cite{Z2007t} and suggest the following discrete symmetry:
\begin{eqnarray}
 U & \rightarrow & U e^{i\pi N}, \nonumber\\
 \theta & \rightarrow & \theta +  \pi N, \nonumber\\
 \Gamma & \rightarrow & \Gamma e^{(2\pi i/3)N},\nonumber\\
 \Pi_4 & \rightarrow & \Pi_4 e^{(2\pi i/4)N},\nonumber\\
 \Pi_5 & \rightarrow & \Pi_5 e^{(2\pi i/5)N}, \nonumber\\
 \Pi_6 & \rightarrow & \Pi_6 e^{(2\pi i/6)N}, \nonumber\\
 ...
\label{symlatWS}
\end{eqnarray}
Here $\Pi_K$ is the $SU(K)$ parallel transporter. We construct our model in
such a way that the parallel transporters correspondent to the new fermions of
the theory are invariant under (\ref{symlatWS}). The resulting symmetry is
denoted by $\cal Z$ and enters expression (\ref{G}).

Let us also point out how our model can be embedded, in principle, into a ETC
model. Let $U(N_{ETC}), N_{ETC} \rightarrow \infty$ be the Unified gauge group.
The breakdown pattern is $U(N_{ETC})\rightarrow ...\otimes SU(5)\otimes
SU(4)\otimes SU(3)\times SU(2) \times U(1)/{\cal Z}$. We may suppose, for
example, that at low energies the $U(N_{ETC})$ parallel transporter has the
form:

\begin{equation}
\Omega = \left( \begin{array}{c c c c c c c}
e^{-2i\theta} & 0 & 0 & 0 & 0 &0& 0\\
0 & Ue^{-i\theta} & 0 & 0 & 0&  0 & 0\\
0 & 0 & \Gamma e^{-\frac{2i\theta}{3}} & 0 & 0 & 0 & 0 \\
0 & 0 & 0 & \Pi_4 e^{-\frac{2i\theta}{4}} & 0 & 0 & 0\\
0 & 0 & 0 & 0 & \Pi_5 e^{-\frac{2i\theta}{5}} & 0 & 0 \\
0 & 0 & 0 & 0 &  0 & \Pi_6 e^{-\frac{2i\theta}{6}} & 0 \\
0 & 0& 0& 0 & 0 & 0 & ...
\end{array}\right)\in U(N_{ETC})
\end{equation}

The form of this parallel transporter demonstrates naturally that the symmetry
(\ref{symlatWS}) is indeed preserved. The fermions of each generation
$\Psi^{i_1 ... i_N}_{j_1 ... j_K}$ carry indices $i_k$ of the fundamental
representation of $U(N_{ETC})$ and the indices $j_k$ of the conjugate
representation. They may be identified with the Standard Model fermions and
Farhi - Susskind fermions as follows (we consider here the first generation
only):

\begin{eqnarray}
&& \Psi^{1} = e_R; \, \Psi_{1}^{1} = \nu_R; \, \Psi^{i_2} =
\left(\begin{array}{c} \nu_L
\\ e^-_L\end{array}\right) ;\nonumber\\
&&\Psi^{i_3} = d_{i_3,R} ; \, \Psi^{i_3}_{1} = u_{i_3,R} ;\, \Psi^{i_2 i_3}_1 =
\left(\begin{array}{c} u^{i_3}_L \\ d^{i_3}_L \end{array}\right) ;\nonumber\\
&&\Psi^{i_4} = E_{i_4,R} ; \, \Psi_{1}^{i_4} = N_{i_4,R} ;\, \Psi^{i_2 i_4}_1 =
\left(\begin{array}{c} N^{i_4}_L \\ E^{i_4}_L \end{array}\right) ;\nonumber\\
&&\Psi^{i_3 i_4} = D_{i_3 i_4,R} ; \, \Psi_1^{i_3 i_4} = U_{i_3
i_4,R} ;\, \Psi^{i_2 i_3 i_4}_{1} = \left(\begin{array}{c} U^{i_3 i_4}_L \\
D^{i_3 i_4}_L
\end{array}\right) \nonumber\\&& (i_2 = 1,3; \, i_3 =
4,5,6;\, i_4 = 7,8,9,10);\label{ferm}
\end{eqnarray}

The other fermions of our model can be arranged in the representations of
$U(N_{ETC})$ in a similar way. From (\ref{ferm}) it follows that all Standard
model fermions can be transformed into the technifermions with ejecting of the
$U(N_{ETC})$ gauge bosons. This is necessary for them to acquire masses. Of
course, the given ETC scheme does not describe the appearance of the realistic
masses for the known particles. However, it gives an example of how, in
principle, the model given in the present paper may be incorporated with the
Extended Technicolor.  We omit here the details of the ETC symmetry breakdown.
We do not describe how does this breakdown occur and what mechanism washes out
the unnecessary fermions. We also do not consider the anomalies in the given
ETC model. We consider all these issues to be out of the scope of the present
paper.

\section{$SU(N)$ groups with $N>4$}

The next step of our investigation is the analysis of the sequence (\ref{F}) in
the form (\ref{i}). Let us notice that the second two rows are actually the
copy of the first two rows supplemented by an additional $SU(3)$ index. Next,
the second four rows are again the copy of the first four rows supplemented by
an additional $SU(4)$ index. Let us suppose that this process is repeated
infinitely. Then the sequence of fermions has the form:
\begin{eqnarray}
... \nonumber\\
U(1), SU(5): && \Psi^{i_5}_{Y_5}, \Psi^{i_5}_{Y_5^\prime};\nonumber\\
U(1), SU(2), SU(5): && \Theta^{i_5 i_2}_{Y_{52}}; \nonumber\\
U(1), SU(3), SU(5): && \Psi^{i_5 i_3}_{Y_{53}}; \Psi^{i_5 i_3}_{Y_{53}^{\prime}}; \nonumber\\
U(1), SU(2), SU(3), SU(5): && \Theta^{i_5 i_3 i_2}_{Y_{532}};\nonumber\\
U(1), SU(4), SU(5): && \Psi^{i_5 i_4}_{Y_{54}},\Psi^{i_5 i_4}_{Y_{54}^{\prime}};\nonumber\\
U(1), SU(2), SU(4), SU(5): && \Theta^{i_5 i_4 i_2}_{Y_{542}};\nonumber\\
U(1), SU(3), SU(4), SU(5): && \Psi^{i_5 i_4 i_3}_{Y_{543}}, \Psi^{i_5 i_4 i_3}_{Y_{543}^\prime};\nonumber\\
U(1), SU(2),SU(3),SU(4), SU(5): &&  \Theta^{i_5 i_4 i_3 i_2}_{Y_{5432}}; \nonumber\\
&&... \nonumber\\
U(1), ... , SU(K): && \Psi^{i_K ... }_{Y_{K...}}, \Psi^{i_K ...}_{Y_{K ...}^\prime};\nonumber\\
U(1), SU(2), ... , SU(K): &&  \Theta^{i_K ... i_2}_{Y_{K...2}}; \nonumber\\
... \label{SMFS100}
\end{eqnarray}

Below we derive the hypercharge assignment for all fermions of our model. We
require that the chiral anomaly is absent and the additional $\cal Z$ symmetry
is preserved. Actually, the fact that there exists such a solution is
 nontrivial. A priory it is not clear that it is possible to satisfy both
 requirements simultaneously.

The anomaly cancellation is always necessary for the model to be well defined.
At the same time the requirement that the $\cal Z$ symmetry is preserved  must
be considered as additional. Of course, at the present moment we do not have
any reason to impose this symmetry but the intuition. So, our reason to
consider the extension of the $Z_6$ symmetry of the Standard Model is the
supposition that it does not appear accidentally. That's why we suppose that it
is to be the manifestation of a more general symmetry. Our choice of Z here is
only one of the possible ways to generalize the $Z_6$. (We hope, however, that
this is one of the most natural ways.) Below it will be shown that $\cal Z$
symmetry gives an important limitation on the choice of fermion hypercharges
and almost fix them (up to the set of integers). Besides, it was shown in
\cite{Z2007t} that the additional discrete symmetry has an important
consequence in the monopole pattern of the Unified model.

Now we require that the chiral anomaly is absent while the gauge group is
(\ref{G}), where $\cal Z$ is defined by (\ref{symlatWS}). Below we prove that
{\bf the necessary hypercharge assignment is}
\begin{eqnarray}
&& Y_2  =  -1 \nonumber\\
&&Y_{i_1 i_2 i_3 ... i_{M-1} i_M 2} = -1 + 2(1 - \frac{1}{i_M})  +  2 \sum_{k =
1}^{M-1}[\theta(i_k - i_{k+1} - 1) - \frac{1}{i_k}] + 2 N_{i_1 i_2 i_3 ...
i_{M-1} i_M 2}
\,\nonumber\\
&&Y_{ i j ... l} = Y_{ i j ... l 2} + 1; \, Y^\prime_{ i j ... l }  = Y_{ i j
... l 2} - 1\label{Y0}
\end{eqnarray}
where $\theta(x) = 1 \, {\rm for} \,x>0; \, \theta(x) = 0 \, {\rm for} \,x\le
0$. In the second row $\, M \ge 1$. For any $K$ integer numbers $N_{i_1 i_2 i_3
... i_{M-1} i_M 2}$ entering (\ref{Y0}) must satisfy the equation
\begin{equation}
\sum_{K > i > j > ... > l > 2} i j ... l \, N_{K ij...l2} = 0 \label{Y11}
\end{equation}
Here the sum is over any (unordered) sets of different integer numbers
$i,j,...,l$ such that $2<i,j,...,l <K$.

{\bf The proof} is as follows. First of all, if (\ref{symlatWS}) is the
symmetry of the theory then the recursion relations take place:
\begin{equation}
Y_{K i j ... l 2} = Y_{i j ... l 2} - \frac{2}{K} + 2 M_{K ij...l2}; Y_{K i j
... l} = Y_{K i j ... l 2} + 1; Y^\prime_{K i j ... l }  = Y_{K i j ... l 2} -
1,
\end{equation}
where $M_{K ij...l2}$ is an integer number.

Let us require that for any $K$
\begin{equation}
\sum_{K > i > j > ... > l > 2} i j ... l \, Y_{K i j ... l 2} = 0,\label{Y}
\end{equation}
 This means that the chiral anomaly is absent  even if the
sequence (\ref{G}) is ended at the $SU(K)$ factor with any value of $K$.

Namely. there may appear the new anomalies of the following types
\cite{Weinberg}:
\begin{eqnarray}
&&1) SU(N) - SU(N) - SU(N),\, N > 2 \nonumber\\
&&2) SU(N) - SU(N) - U(1),\, N>2 \nonumber\\
&&3) SU(2) - SU(2) - U(1) \nonumber\\
&&4) U(1) - U(1) - U(1)
\end{eqnarray}

The anomaly of the first type vanishes because the number of left - handed
fermions is equal to the number of the right - handed ones while both types of
fermions belong to the fundamental representation of $SU(N)$. The anomalies of
the second type vanish because $Y_{ i j ... l} = Y_{ i j ... l 2} + 1; \,
Y^\prime_{ i j ... l }  = Y_{ i j ... l 2} - 1$. The anomalies of the third and
the fourth types vanish if the sum of the hypercharge over left - handed
doublets is zero. This leads to (\ref{Y}).

Below we prove that for any $K$ integer numbers $M_{K ij...l2}$ can be chosen
in such a way, that (\ref{Y}) is satisfied. Let $\sum_{K^\prime > i > j > ... >
l > 2} i j ... l Y_{K^\prime i j ... l 2} = 0$ for $K^\prime < K$ (this was
demonstrated already for $K^\prime = 4$.). Then
\begin{equation}
\sum_{K > i > j > ... > l > 2} i j ... l \, Y_{K i j ... l 2} =  - 2
\frac{K!}{3!K} + 2\sum_{K > i > j
> ... > l > 2} i j ... l \, M_{K ij...l2} \label{KM}
\end{equation}
Here we used the identity $\sum_{K > i > j > ... > l > 2} i j ... l =
\frac{K!}{3!}$.

Let us know introduce the following notations:
\begin{eqnarray}
M_{K ij...l2} = M_{K ij...l2}^\prime + 1,\, {\rm for}\, \, K-1 > i > j > ... >
l > 2;\nonumber\\
M_{K ij...l2} = M_{K ij...l2}^\prime,\, {\rm for}\, \, K-1 = i > j > ... > l >
2
\end{eqnarray}

Then
\begin{equation}
-\frac{K!}{3!K} + \sum_{K > i > j > ... > l > 2} i j ... l \, M_{K ij...l2}  =
\sum_{K > i > j > ... > l > 2} i j ... l \, M^\prime_{K ij...l2}
\end{equation}

The relations that define the fermion hypercharges can be rewritten in the
following way:
\begin{eqnarray}
&&Y_{K i j ... l} = Y_{K i j ... l 2} + 1; \, Y^\prime_{K i j ... l }  = Y_{K i
j ... l 2} - 1; Y_{K i j ... l 2} = Y_{i j ... l 2} - \frac{2}{K} + 2 + 2
M^\prime_{K ij...l2} \,\nonumber\\ && ( {\rm for}\, \, K-1
> i > j > ... > l > 2, \, {\rm or} \, K = 3);\nonumber\\
&&Y_{K i j ... l 2} = Y_{i j ... l 2} - \frac{2}{K}  + 2 M^\prime_{K ij...l2}
\, ({\rm for}\, \, K-1 = i > j > ... > l > 2)
\end{eqnarray}
Here integer numbers $M^\prime_{K ij...l2}$ are chosen in such a way that
$\sum_{K > i > j > ... > l > 2} i j ... l \, M^\prime_{K ij...l2} = 0$.

Finally we come to the solution of (\ref{Y}) in the form (\ref{Y0}). In
particular, the choice $N_{i_1 i_2 i_3 ... i_{M-1} i_M 2}=0$ corresponds to
$Y_{i_1 i_2 i_3 ... i_{M-1} i_M 2} = -1 + 2(1 - \frac{1}{i_M}) +  2 \sum_{k =
1}^{M-1}[\theta(i_k - i_{k+1} - 1) - \frac{1}{i_k}]$.
 Thus the additional symmetry (\ref{symlatWS}) fixes the hypercharge
assignment up to the choice of integer numbers $N_{i_1 i_2 i_3 ... i_{M-1} i_M
2}$ such that (\ref{Y11})  is satisfied.

It is worth mentioning that if the $\cal Z$ symmetry is not imposed, then the
hypercharge assignment is defined by the anomaly cancellation only. In this
case the hypercharge assignment is given by (\ref{Y0}), where $N_{i_1 i_2 i_3
... i_{M-1} i_M 2}$  are not necessarily integer.

\section{Discussion}

The dynamics of Technicolor is related in a usual way to the number of fermions
$N_f$. Namely, the beta - function in one loop approximation has the form: $
\beta_{SU(K)}(\alpha) = - \frac{11 K - 2 N_f}{6 \pi} \alpha^2$ where $\alpha =
\frac{g^2_{SU(K)}}{4\pi}$. If $N_f < \frac{11}{2} K$, the one loop calculation
indicates asymptotic freedom. The two - loop calculations
\cite{Appelquist,Appelquist_} indicate that the chiral symmetry breaking occurs
at $N_f < N_c \sim K \frac{100 K^2 - 66}{25 K^2 - 15} \sim 4 K $. This is
required for the appearance of gauge boson masses.

In our model we have  three generations of Farhi - Susskind technifermions.
Therefore, their number is $24 > 4 N_{TC} = 16$. However, it is important that
only such technifermions are relevant, the masses of which are of the order of
$\Lambda_{TC}$ and smaller ($\Lambda_{TC}$ is the $SU(4)$ analogue of
$\Lambda_{QCD}$). Therefore, we suppose that the masses of the third generation
technifermions and, probably, the masses of some of the second generation
technifermions  are essentially larger, than the Technicolor scale. We also
assume that the masses of the fermions that carry the indices of higher
Hypercolor groups are essentially larger than the Technicolor scale. So, they
do not affect the Technicolor dynamics. Thus  the $SU(4)$ interactions lead to
the chiral symmetry breaking and provide $W$ and $Z$ bosons with their masses.

If the number of fermions approach $N_c \sim 4 N_{TC}$, then the behavior of
the model becomes close to conformal. In this case the effective charge becomes
walking instead of running \cite{walking}. So, in our case (two generations of
fermions for $N_{TC} = 4$) the behavior of the technicolor may be close to
conformal.

As for the higher Hypercolor groups, already for $SU(5)$ interactions the
number of the first generation hyperfermions (fermions carrying $SU(5)$ index)
is $2(1 + 3 + 4 + 12) = 40 > \frac{55}{2} = 27.5$. We suppose their masses are
close to each other. That's why the Hypercolor forces at $K > 4$ are not
asymptotic free, and do not confine. As a result the Landau pole is present in
their effective charges. This means that our model does not have a rigorous
continuum limit, and should be considered as a finite cutoff model. At the
energies of the order of this cutoff the new theory should appear that
incorporates the Hypercolor tower as an effective low energy theory. In
principle, this scale may be extremely large, even of the order of Plank mass
depending on the value of $g^2_{SU(K)}$ at low energies. Very roughly this
scale (as given by the $SU(5)$ effective charge) can  be estimated as
$\Lambda_h = e^{\frac{6 \pi}{(2N_f - 55) \alpha_{SU(5)}(1 \, {\rm Tev})}} \,
{\rm Tev}$. Say, if three generations are involved, and $\alpha_{SU(5)}(1 \,
{\rm Tev}) = \frac{1}{300}$, then the Landau pole occurs in the $SU(5)$ gauge
coupling at $\Lambda_h \sim 10^{13}\, {\rm Tev}\sim 10^{-3} M_{\rm Plank}$. At
the energies much less than $\Lambda_h$ the $SU(5)$ interactions can be taken
into account perturbatively just like in QED. However, the description of
possible effects due to Hypercolor $SU(5)$ (and due to the other Hypercolor
interactions $SU(N)$ for $N>4$) is out of the scope of the present paper.

This work was partly supported by RFBR grants  09-02-00338, 08-02-00661, and
07-02-00237, by Grant for leading scientific schools 679.2008.2, by Federal
Program of the Russian Ministry of Industry, Science and Technology No
40.052.1.1.1112.

\clearpage

\end{document}